\documentclass[table]{IEEEtran}
\usepackage[utf8]{inputenc}
\usepackage{array}
\usepackage{wrapfig}
\usepackage{multirow}
\usepackage{tabu}
\usepackage{xcolor}
\usepackage{cite}
\usepackage{amsmath,amssymb,amsfonts}
\usepackage{algorithmic}
\usepackage{graphicx}
\usepackage{textcomp}
\usepackage{caption}
\usepackage{comment}
\usepackage{subcaption}
\usepackage{float}
\usepackage{eucal}
\usepackage{soul}
\usepackage{fixltx2e}
\usepackage{enumerate}
\usepackage{cuted}
\usepackage{pgf-pie}
\def\BibTeX{{\rm B\kern-.05em{\sc i\kern-.025em b}\kern-.08em
    T\kern-.1667em\lower.7ex\hbox{E}\kern-.125emX}}

\usepackage{fancyhdr}

\fancypagestyle{plain}{
  \fancyhf{}
  \fancyhead[C]{}     
  \fancyfoot[L]{This paper has been accepted at the IEEE International Symposium on Circuits and Systems (ISCAS) to be held in May 2020 at Seville, Spain.}

}
\usepackage{eso-pic}

\begin{document}

\AddToShipoutPictureBG*{%
  \AtPageLowerLeft{%
    \setlength\unitlength{1in}%
    \hspace*{\dimexpr0.5\paperwidth\relax}
    \makebox(0,0.75)[c]{\small This paper has been accepted at the IEEE International Symposium on Circuits and Systems (ISCAS) to be held in May 2020 at Seville, Spain.}%
}}

\title{MC$^2$RAM: Markov Chain Monte Carlo Sampling in SRAM for Fast Bayesian Inference}
\author{Priyesh Shukla, Ahish Shylendra, Theja Tulabandhula, and Amit Ranjan Trivedi \\
University of Illinois at Chicago, IL, USA, email: \{pshukl23, amitrt\}@uic.edu
\vspace{-1.75em}}

\maketitle
\vspace{-2.5em}
\begin{abstract}
This work discusses the implementation of Markov Chain Monte Carlo (MCMC) sampling from an arbitrary Gaussian mixture model (GMM) within SRAM. We show a novel architecture of SRAM by embedding it with random number generators (RNGs), digital-to-analog converters (DACs), and analog-to-digital converters (ADCs) so that SRAM arrays can be used for high performance Metropolis-Hastings (MH) algorithm-based MCMC sampling. Most of the expensive computations are performed within the SRAM and can be parallelized for high speed sampling. Our iterative compute flow minimizes data movement during sampling. We characterize power-performance trade-off of our design by simulating on 45 nm CMOS technology. For a two-dimensional, two mixture GMM, the implementation consumes $\sim91\mu\/W$ power per sampling iteration and produces 500 samples in 2000 clock cycles on an average at 1 GHz clock frequency. Our study highlights interesting insights on how low-level hardware non-idealities can affect high-level sampling characteristics, and recommends ways to optimally operate SRAM within area/power constraints for high performance sampling.

\end{abstract}

\begin{IEEEkeywords}
Inference; in-memory computing; Markov chain Monte Carlo (MCMC) sampling.
\vspace{-0.5em}
\end{IEEEkeywords}

\section{Introduction}
\label{sec:introduction}
\begin{figure}[t]
    \centering
    \begin{subfigure}[!t]{\linewidth}
         \centerline{\includegraphics[width=\columnwidth]{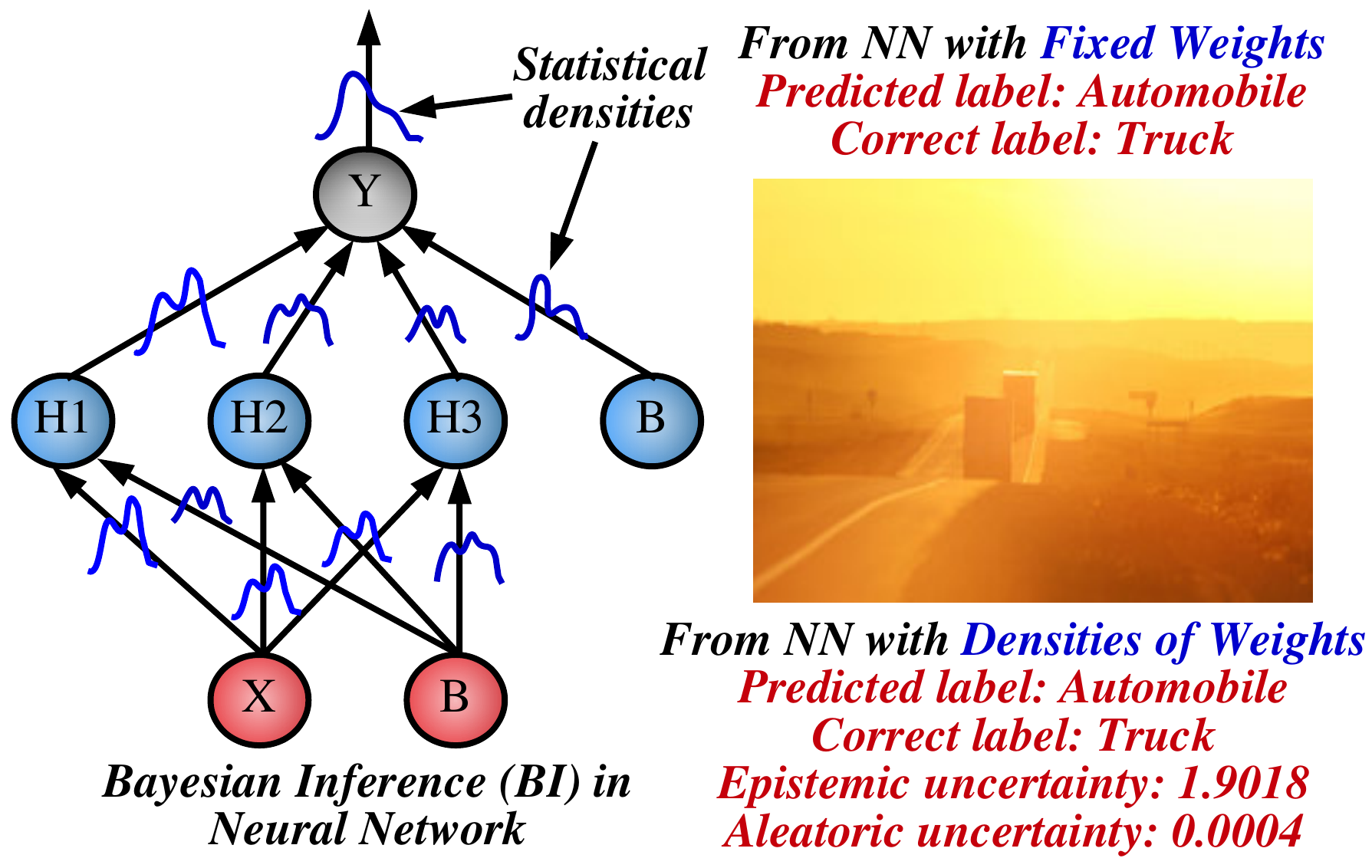}}
        \end{subfigure}
    \caption{\small{(Left) Bayesian inference (BI) in neural network (NN) represents weights as a density function, unlike classical inference (CI), where weights assume scalar values. (Right) An ambiguous image of trucks in glaring sunlight \cite{kyle_dorman} causing uncertainty in prediction and decision making. BI accounts for prediction-uncertainty unlike classical inference.}}
    \label{fig:BI_images}
    \vspace{-1.35em}
\end{figure}

Markov chain Monte Carlo (MCMC) is an extensively used statistical sampling technique for generating samples from high-dimensional probability density functions even when these functions can not be defined analytically \cite{Andrieu03anintroduction, neal_stats_doi:10.1080/10618600.2000.10474879}. Especially, in recent years, as various machine learning (ML) platforms are proliferating for realtime decision-making, MCMC is being combined with Bayesian ML models to perform efficient inference \cite{Hinton1995BayesianLF, GhavamzadehMAL-049}. Unlike classical inference, Bayesian inference can capture uncertainties in the outcomes for risk-aware decision making \cite{blundell2015weight} as shown in Fig. \ref{fig:BI_images}. Moreover, there is a growing interest to operate ML-based prediction models at the edge itself \cite{horowitz6757323, sze8114708}. A low power/area MCMC platform is, therefore, becoming imperative along with low power ML implementation. 

Prior works have discussed low power MCMC implementations using FPGAs, and have achieved an accuracy similar to their software counterparts, while being more energy-efficient. An FPGA-based hardware accelerator \cite{Cai:2018:VHA:3173162.3173212} was designed for variational inference of Bayesian neural networks (BNNs). Conversely, in this work, we present MC$^2$RAM -- a customized implementation of MCMC within SRAM where we co-locate and co-optimize functional units, control flow, and data flow to address critical bottlenecks in high-speed MCMC-based sampling. Our compute flow exploits the Markov chain property where successive chain outputs lie in the proximity minimizing the necessary computing load in each iteration. While MCMC in prior works \cite{Cai:2018:VHA:3173162.3173212, Banerjee:2019:AAM:3297858.3304019} is limited to Gaussian functions, MC$^2$RAM expands this to Gaussian mixture models (GMMs). A GMM can model any density arbitrarily closely with enough mixture components, making our implementation vastly more applicable. We develop interesting insights about the interaction between low-level hardware non-idealities and high-level sampling characteristics in MC$^2$RAM. Our detailed design and operating power space exploration can lead to efficient design methodologies for SRAM-based sampling.

The paper is organized as follows. Section II discusses the background on MCMC and provides an overview of MC$^2$RAM. Section III discusses density function computation and sampling in MC$^2$RAM. Section IV discusses the results and implications of various sources of non-idealities in MC$^2$RAM on sampling. Section V concludes this paper.
\vspace{-0.25em}

\section{MCMC for BI and proposed MC$^2$RAM}
\label{sec:background}
In ML models and methods that employ Bayesian inference (BI), the expectation of the predicted outcome (or other quantity of interest) is obtained by solving $\int_{}^{}M(I,w) \times P(w|D)dw$, where $M(I, w)$ is the model (say, a neural network) with the input $I$ and parameters/weights $w$, and $P(w|D)$ is the posterior density of weights given training data $D$. In such computations, an analytical integration is often intractable since the density function of the random variable (RV) and/or the function to be integrated, e.g., $P(w|D)$ and $M(I, w)$ in BI, are too complicated. A Monte Carlo approach, therefore, becomes necessary to numerically compute these quantities. Monte Carlo approach reduces an integral over a function of RV, $x$, as

\begin{equation} \label{Eq:BI_eqn}
\int_{}^{}G(x) \times \mathcal{F}(x)dx \approx \frac{1}{T} \times \Sigma_{t=1}^{T} G(x_{\mathcal{F}(x)})
\end{equation}
\begin{figure}[t]
    \centering
    \begin{subfigure}[!t]{\linewidth}
         \centerline{\includegraphics[width=0.775\columnwidth, height=3.5cm]{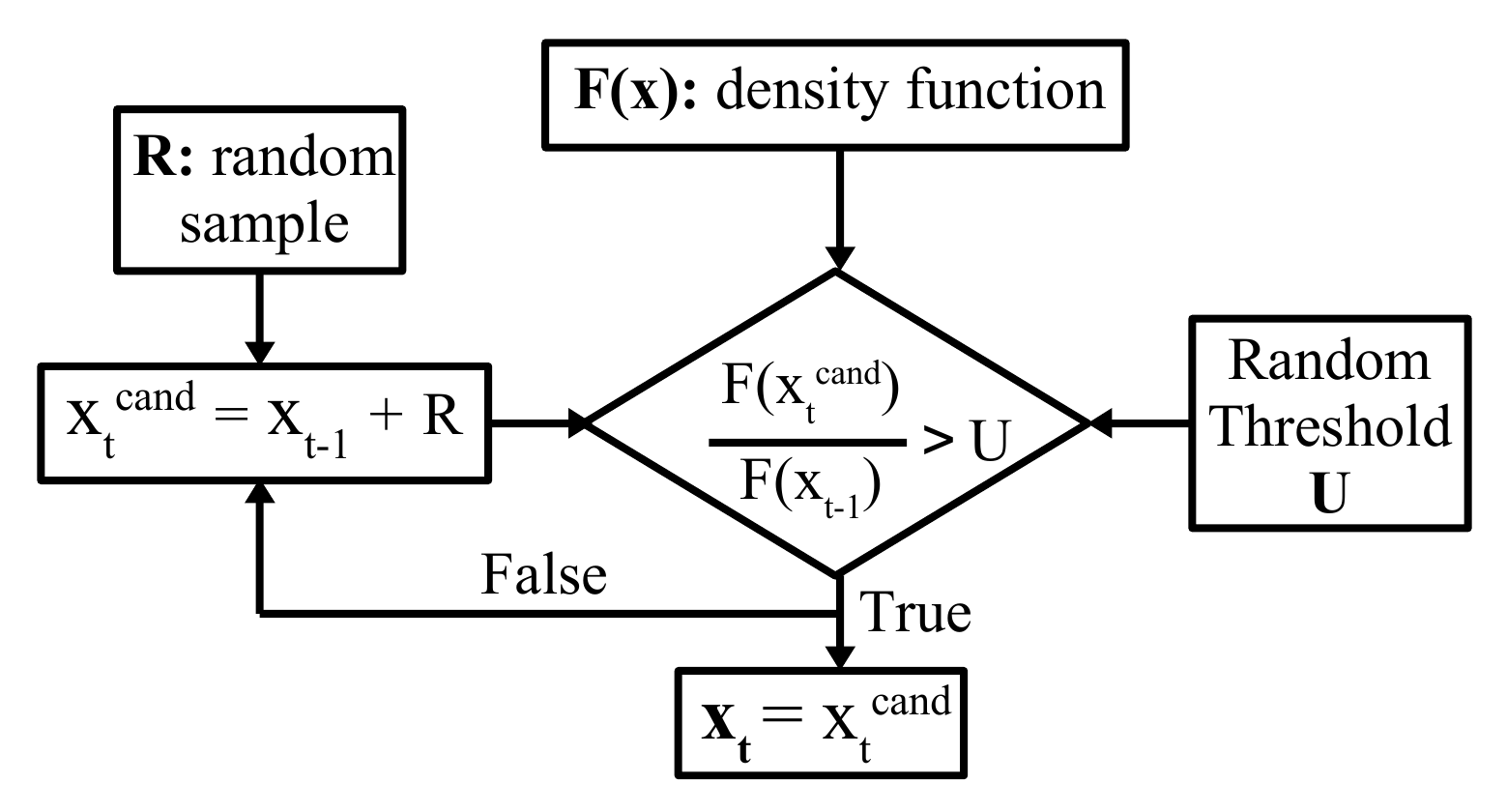}}
        \end{subfigure}
    \caption{\small{System level flow for in-SRAM MCMC sampling.}}
    \label{fig:high_level_flow}
    \vspace{-1.25em}
\end{figure}
Here, $G(x)$ is the function to be integrated, $\mathcal{F}(x)$ is the density function of $x$, and $x_{\mathcal{F}(x)}$ is an independent and identically distributed (i.i.d.) sample drawn from $\mathcal{F}(x)$. The law of large numbers guarantees an asymptotic convergence of the summation to the exact integral as the number of samples ($T$) increases. In BI, $\mathcal{F}(x)$ is the posterior density $P(w|D)$, which can be numerically extracted using the Bayesian formula $P(w|D) \propto P(D|w)\times P(w)$, but cannot always be defined analytically. Therefore, MCMC overcomes this critical problem by defining an ergodic Markov chain. Among the MCMC methods for sampling we choose Metropolis-Hastings (MH) sampling \cite{chib10.2307/2684568} that provides a middle ground in terms of acceptance/rejection complexity of the samples and the average time needed to generate a candidate sample. The algorithmic steps for MH-based MCMC are demonstrated in Fig. \ref{fig:high_level_flow}. A candidate sample at step $t$, $x_{t}^{cand}$ is determined from the previously accepted sample $x_{t-1}$ and a randomly generated sample, $R$, as $x_{t}^{cand}$ = $R$ + $x_{t-1}$; here, $R$ follows the statistics of the proposal distribution $\mathcal{P}$ (e.g., Uniform or Gaussian) typically centered at zero. The statistics of $R$ controls the search radius and sample search behavior in MCMC. The candidate sample is accepted when the ratio of the density of $x_{t}^{cand}$ to that of $x_{t-1}$, i.e., $\mathcal{F}(x_{t}^{cand})/\mathcal{F}(x_{t-1})$ is greater than a random threshold, $U$, which is generated uniformly between zero and one.

\label{sec:background}

\begin{figure}[t]
\centerline{\includegraphics[width=0.85\columnwidth, height=6.30cm]{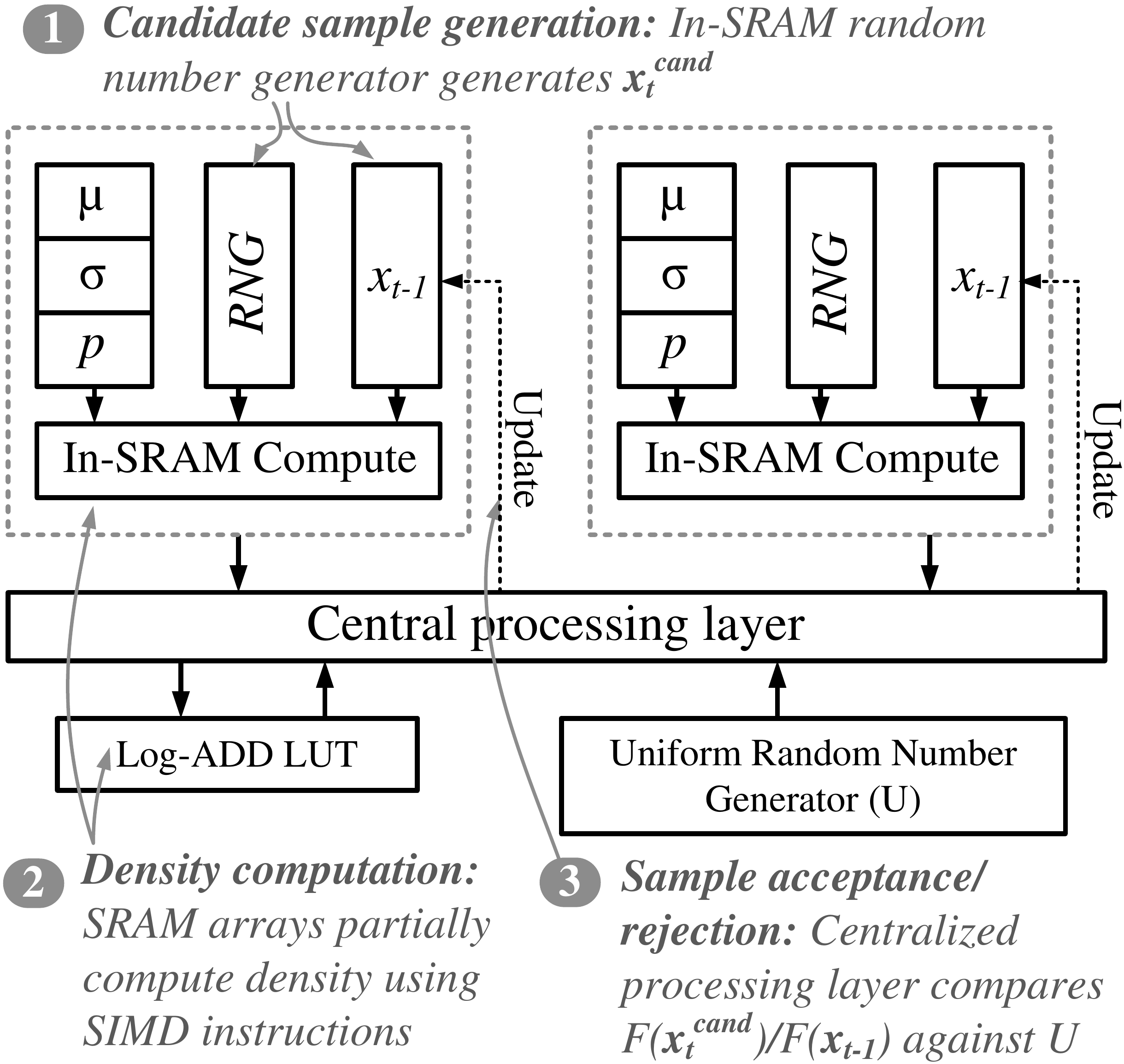}}
\caption{\small{Overview of MC$^2$RAM for within SRAM MCMC sampling.}}
\label{fig:high_level_system}
\vspace{-1.25em}
\end{figure}

Fig. \ref{fig:high_level_system} shows the overall architecture of MC$^2$RAM. An SRAM array stores the GMM parameters for the density function of the RV, i.e, mean, variance, and mixture weights ($\mu, ~\sigma, ~\& ~p$). Since the dimension of the sampling density can be high, $\mu, ~\sigma, ~\& ~p$ vectors are stored appropriately in multiple SRAM banks as shown in the figure. SRAM arrays are also integrated with random number generators (RNG). Using RNGs and a previous sample of the chain ($x_{t-1}$), a candidate sample ($x_t^{cand}$) is generated within the SRAM. For $x_t^{cand}$, SRAM arrays partially compute the density of the candidate sample, i.e., $\mathcal{F}(x_t^{cand})$, in parallel by following single instruction multiple data (SIMD) style execution. Central processing layer receives the partial terms for density computation from SRAM arrays and applies Metropolis-Hastings-based sample acceptance criteria to accept/reject $x_t^{cand}$. Using the accepted $x_t$, the chain iterates to find the next sample $x_{t+1}$. 

The key complexities of MCMC illustrated via Fig. \ref{fig:high_level_flow} are: (i) computations of $\mathcal{F}(x_t^{cand})$ for a candidate sample $x_t^{cand}$ since the dimension of $x_t^{cand}$ can be high and/or density function $\mathcal{F}(x)$ can be complex, and (ii) high throughput sampling since many $x_t^{cand}$ may end up being rejected. The details of our framework addressing these complexities is discussed subsequently.

\begin{figure*}[t]
    \centering
    \begin{subfigure}[!t]{0.38\linewidth}
         \centerline{\includegraphics[width=\linewidth, height=4.5cm]{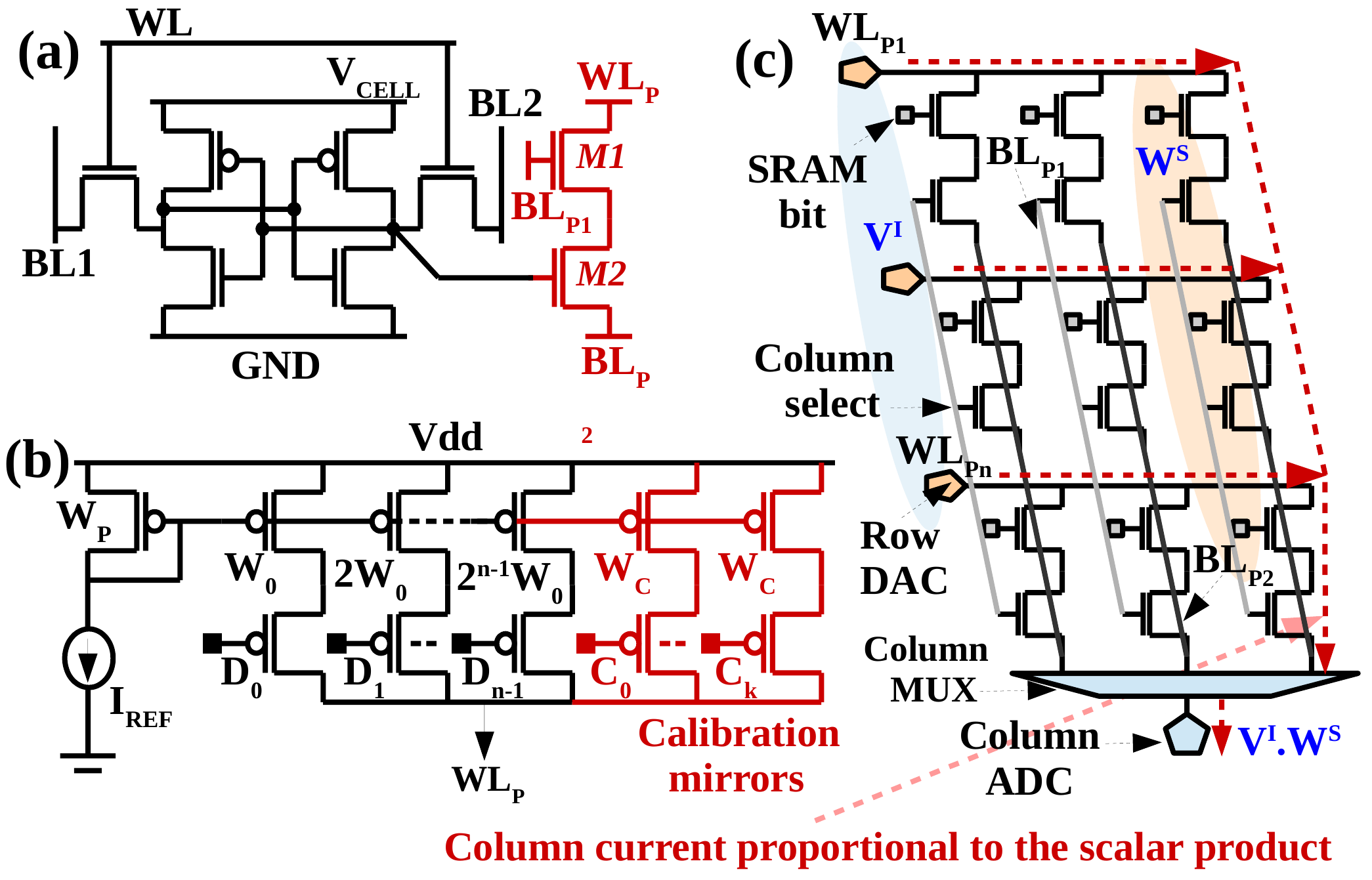}}
        \end{subfigure}
          \begin{subfigure}[!t]{0.30\linewidth}
          \centerline{\includegraphics[width=\linewidth, height=4.5cm]{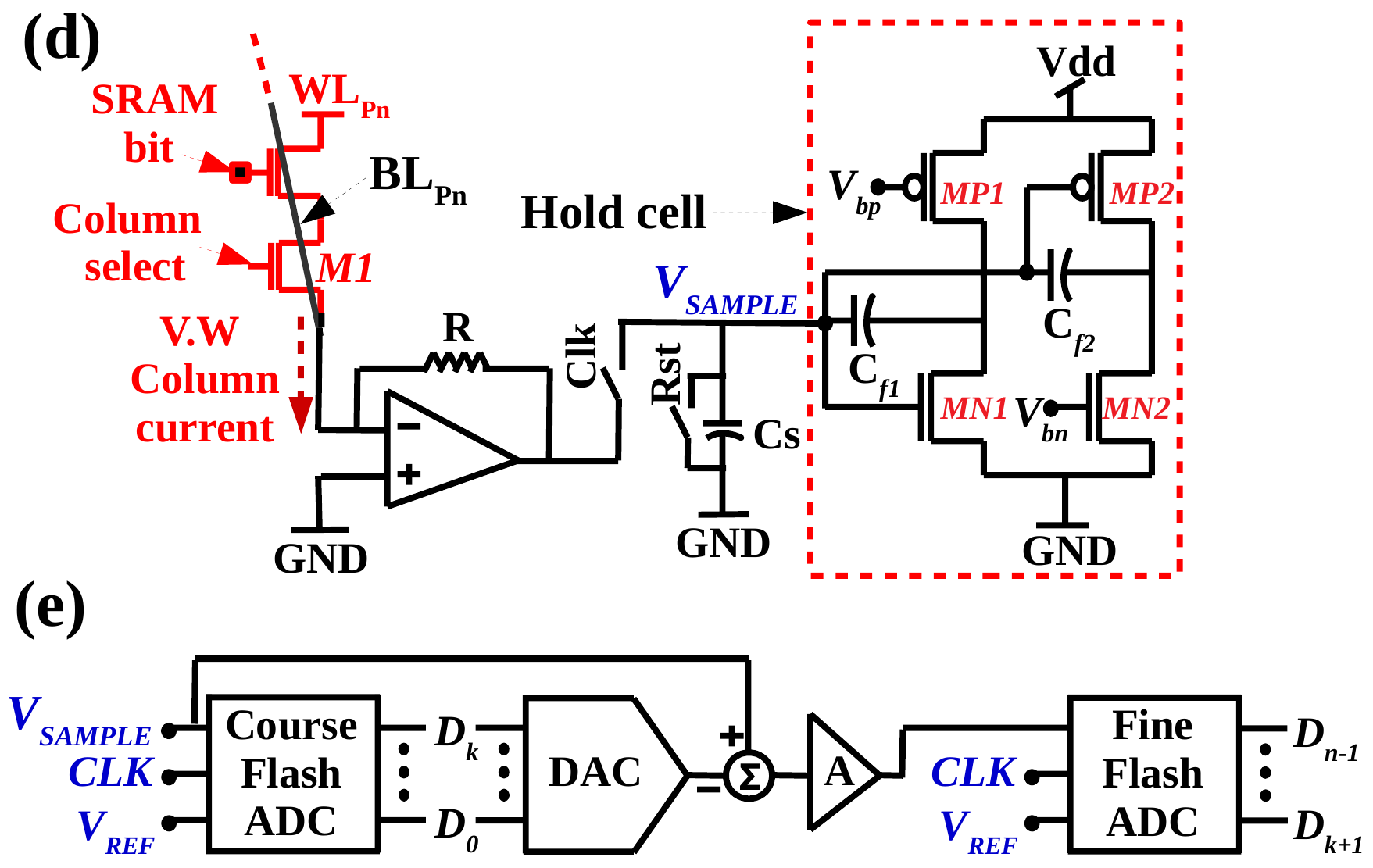}}
        \end{subfigure}
         \begin{subfigure}[!t]{0.27\linewidth}
        \centerline{\includegraphics[width=\linewidth, height=4.5cm]{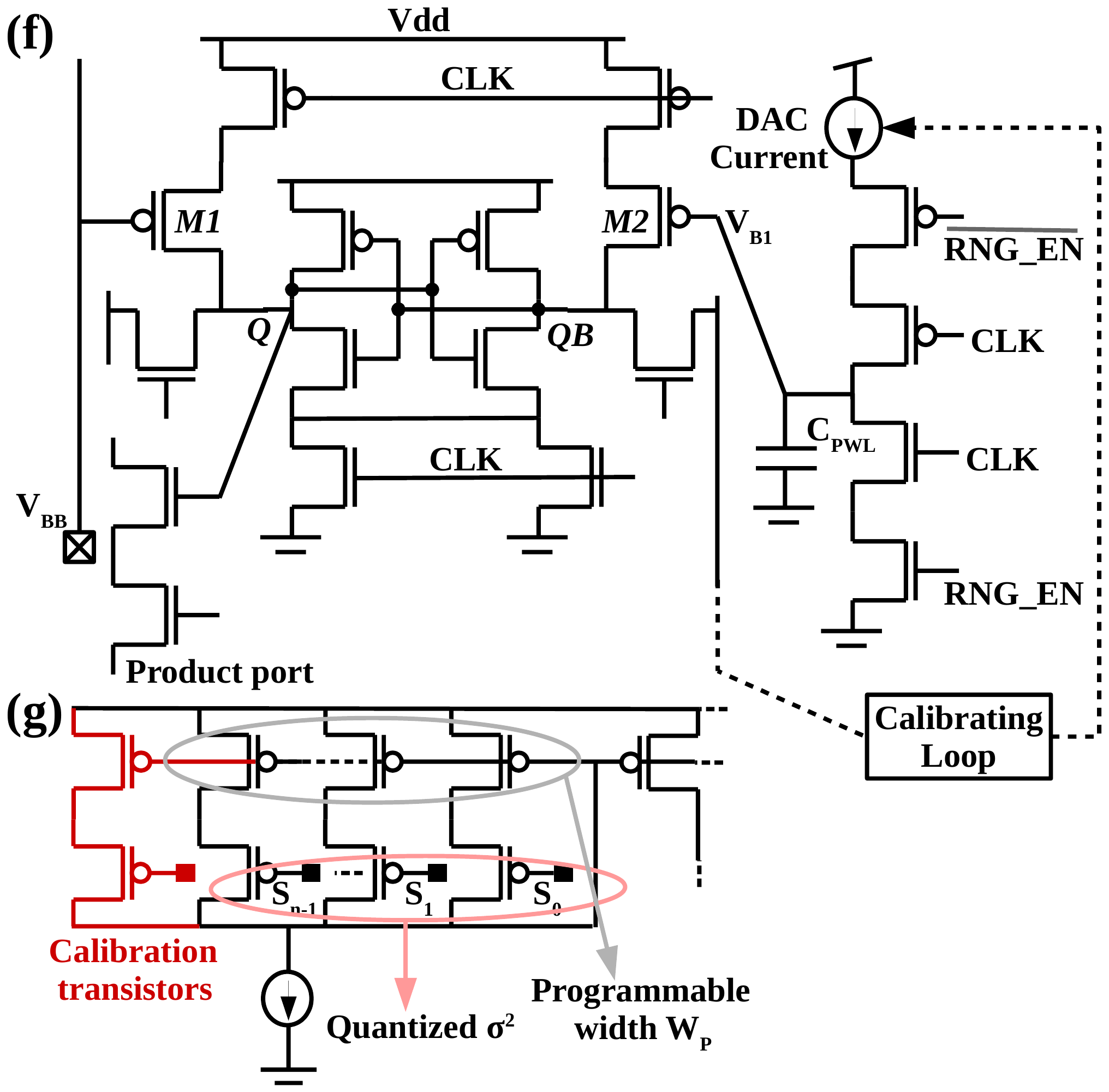}}
        \end{subfigure}
        
    \caption{\small{(a) 8-T SRAM cell with product port. (b) DAC co-located with SRAM cell. (c) Current-mode scalar product operation within SRAM. (d-e) Op-Amp to stabilize column current and interface with ADC. (f-g) In-SRAM RNG cell.}}
    \label{fig:circuits}
    \vspace{-1.25em}
\end{figure*}

\begin{figure}[t]
    \centering
    \begin{subfigure}[!t]{\linewidth}
         \centerline{\includegraphics[width=0.75\columnwidth, height=6cm]{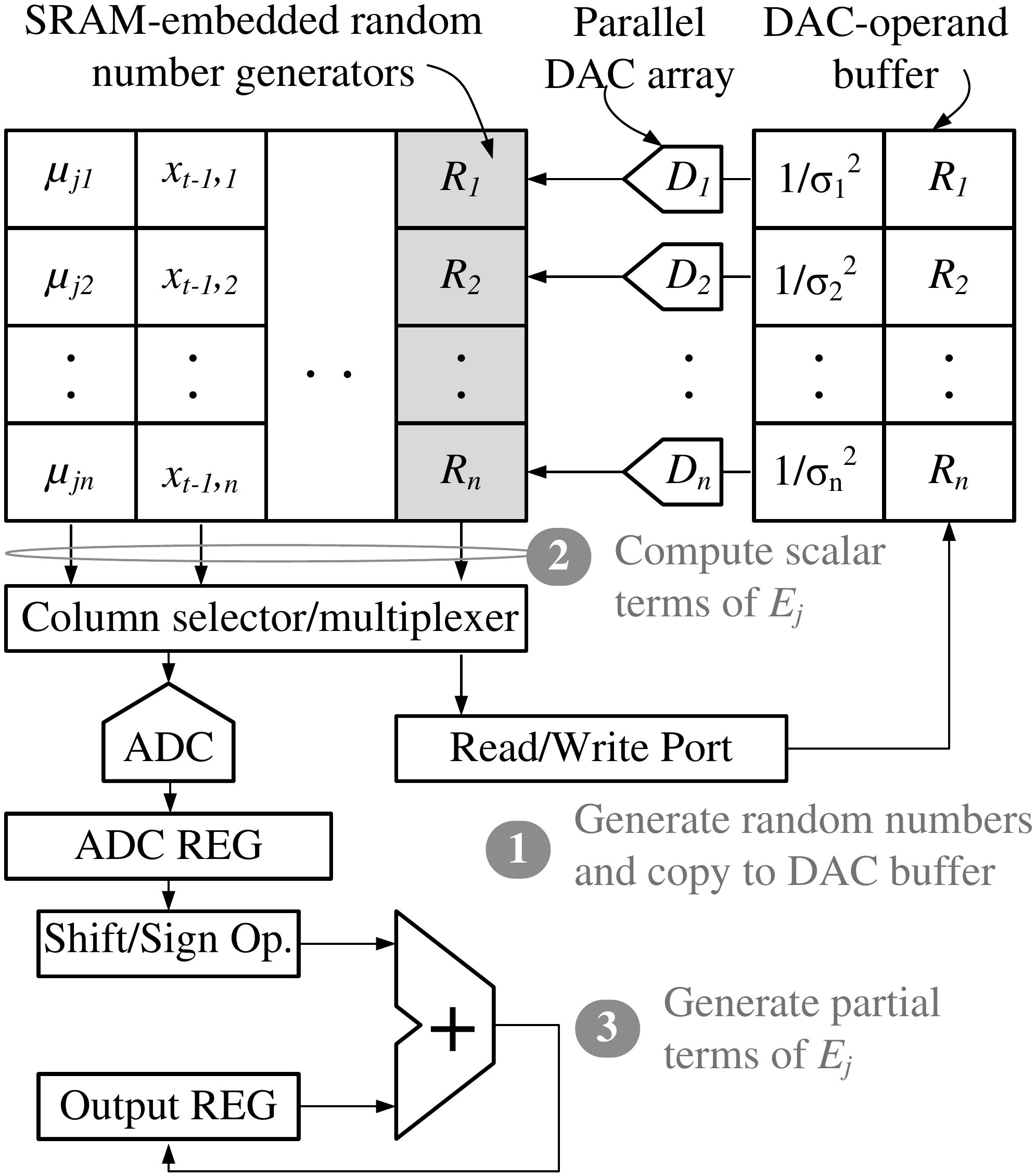}}
        \end{subfigure}
    \caption{\small{System level flow for in-SRAM MCMC sampling.}}
    \label{fig:block_level_Ej}
    \vspace{-1.75em}
\end{figure}

\section{In-SRAM Density Computation and Sampling}

The density of a GMM, $\mathcal{G}(x)$, for a candidate sample $x_t^{cand}$ is given by 
\vspace{-1em}
\begin{equation}
\mathcal{G}(x_t^{cand}) = \sum_{j=1}^{M}p_j \times \mathcal{N}(x_t^{cand};\mu_j,\sigma_j)
\end{equation}
Here, $M$ is the total number of mixture components; $\mathcal{N}(x_t^{cand};\mu_j,\sigma_j)$ is $j^{th}$ Gaussian mixture function whose density depends on $\mu_j$ and $\sigma_j$; and $p_j$ is the mixture weight. In approximating a density function $\mathcal{F}(x)$ using GMM, mixture Gaussians with only diagonal co-variance can be used (this is called as mean-field approximation \cite{cmu.meanfieldapproximation}). The density of $\mathcal{N}(x_t^{cand};\mu_j,\sigma_j)$ depends on its exponent $E_j$ as
\vspace{-0.5em}
\begin{equation}
E_j = \sum_{i=1}^{N} \Bigg(\frac{x^{cand}_{i}-\mu_{ij}}{\sigma_{ij}}\Bigg)^{2}
\end{equation}
Here, $x^{cand}$, $\mu_{j}$ and $1/\sigma_{j}$ are each $N$-dimensional and expanded using the subscript `$i$' as in the above equation. The overall GMM density in log-domain can be computed from the exponents $E_j$ itself using the identity $ln(e^a + e^b) = a + ln(1 +  e^{b-a})$ and a look up table (LUT) for $ln(1+ e^x)$. $E_j$ can be further simplified by exploiting the sampling property where $x^{cand}_t$ is searched in the proximity of the previous MCMC sample $x_{t-1}$. Thus, $E_j(t)$ at $x^{cand}_t$ can be computed from $E_{j}(t-1)$ at $x_{t-1}$ by
\vspace{-0.25em}
\begin{equation}
E_{j}(t) = E_{j}(t-1) + \Big(\frac{R}{\sigma_{j}^{2}} \cdot R\Big) + 2 \cdot (\frac{R}{\sigma_{j}^{2}}) \cdot (x_{t-1}-\mu_{j})
\end{equation}
Here, $R$ is a generated random number from the proposal distribution $\mathcal{P}$ within SRAM, which is used to search the next MCMC sample. $R/\sigma_{j}^{2}$ is an $N$-dimensional vector obtained by dividing each element of $R$, $R_i$, with the corresponding $\sigma_{ij}^{2}$. We apply SRAM to compute scalar products $R \cdot R/\sigma_{j}^{2}$ and $(x_{t-1}-\mu_{j}) \cdot R/\sigma_{j}^{2}$ within SRAM array to evaluate (4).

\begin{figure*}[h]
    \centering
    \begin{subfigure}[!t]{0.19\linewidth}
         \centerline{\includegraphics[width=\linewidth, height=4cm]{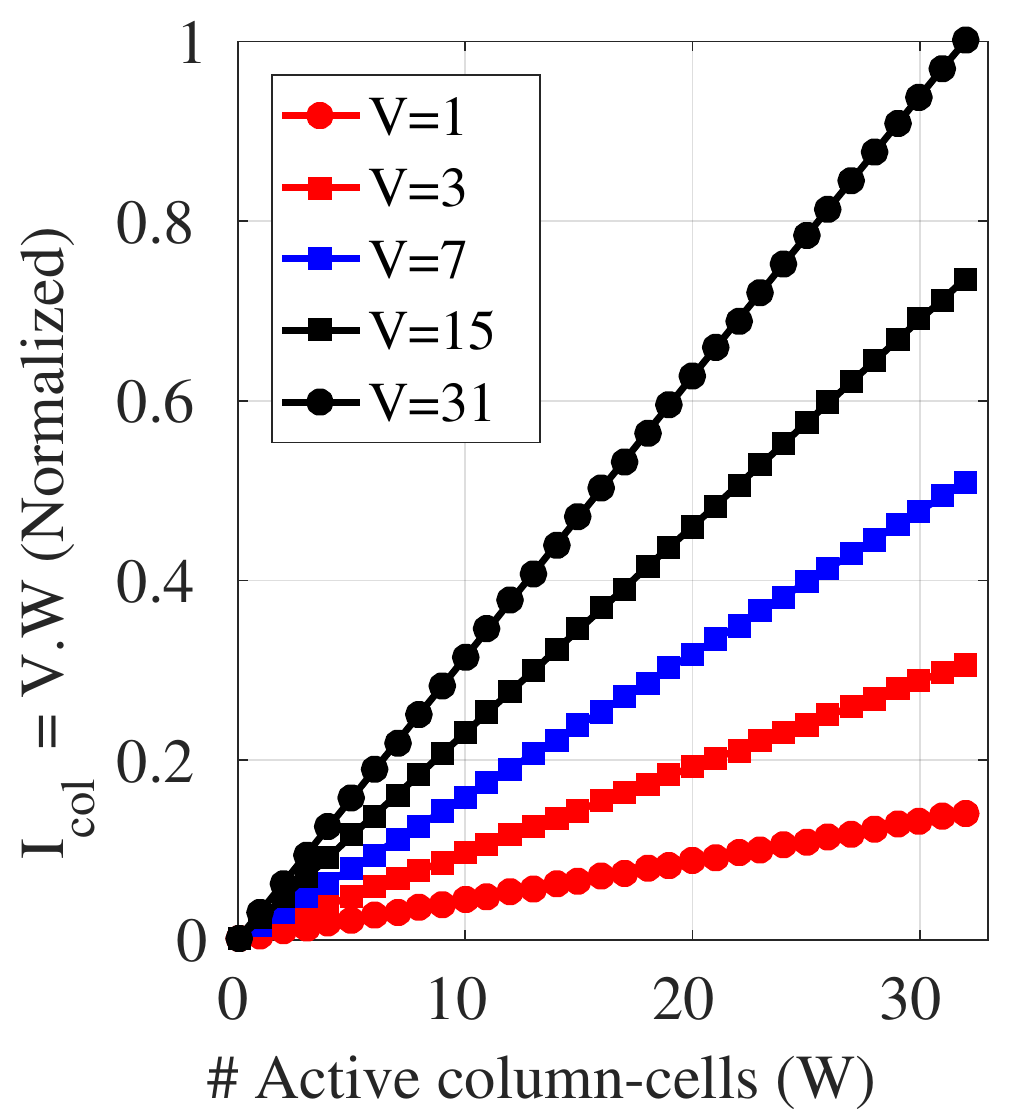}}
         \caption{}
        \end{subfigure}
         \begin{subfigure}[!t]{0.19\linewidth}
        \centerline{\includegraphics[width=\linewidth, height=4cm]{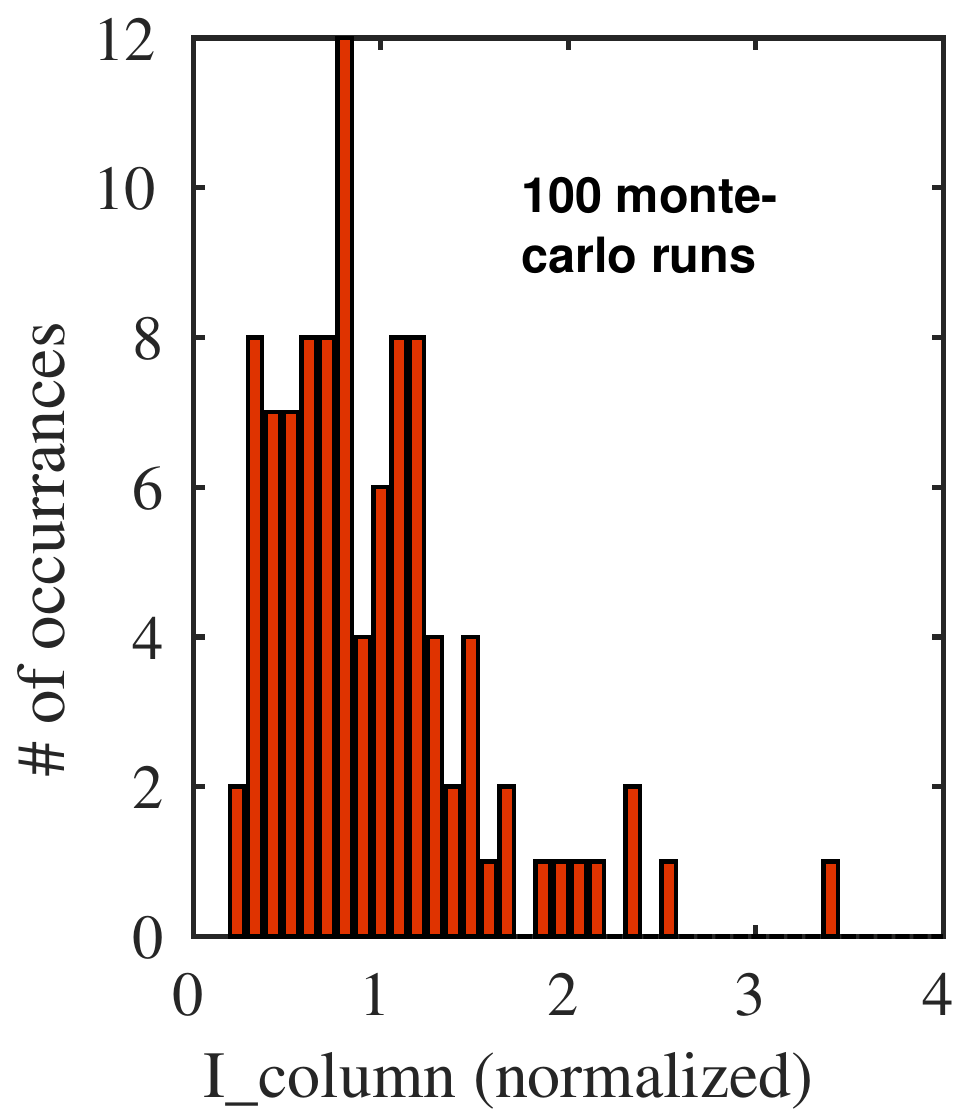}}
           \caption{}
        \end{subfigure}
        \begin{subfigure}[!t]{0.19\linewidth}
          \centerline{\includegraphics[width=\linewidth, height=4cm]{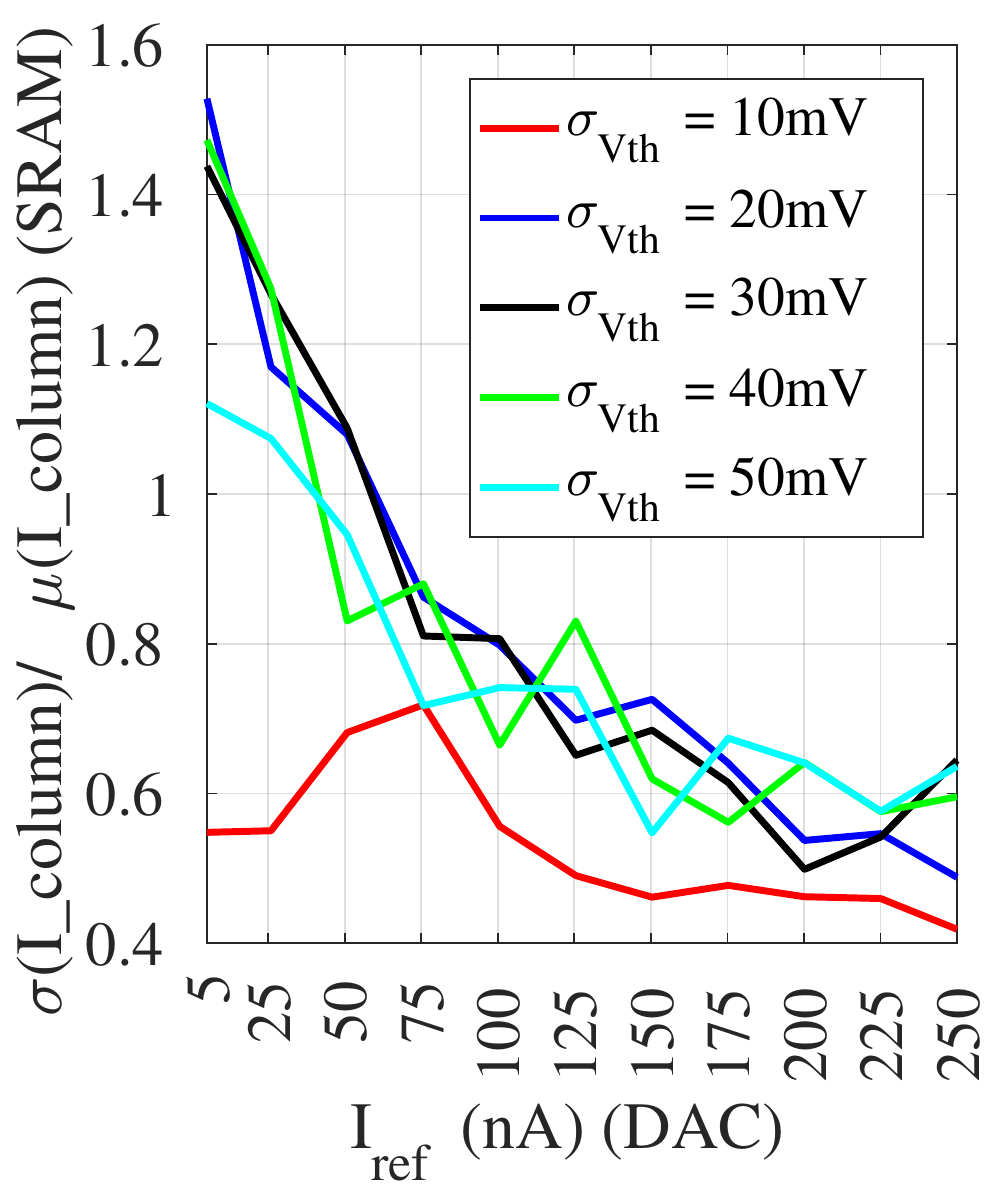}}
           \caption{}
        \end{subfigure}
        \begin{subfigure}[!t]{0.19\linewidth}
        \centerline{\includegraphics[width=\linewidth, height=4cm]{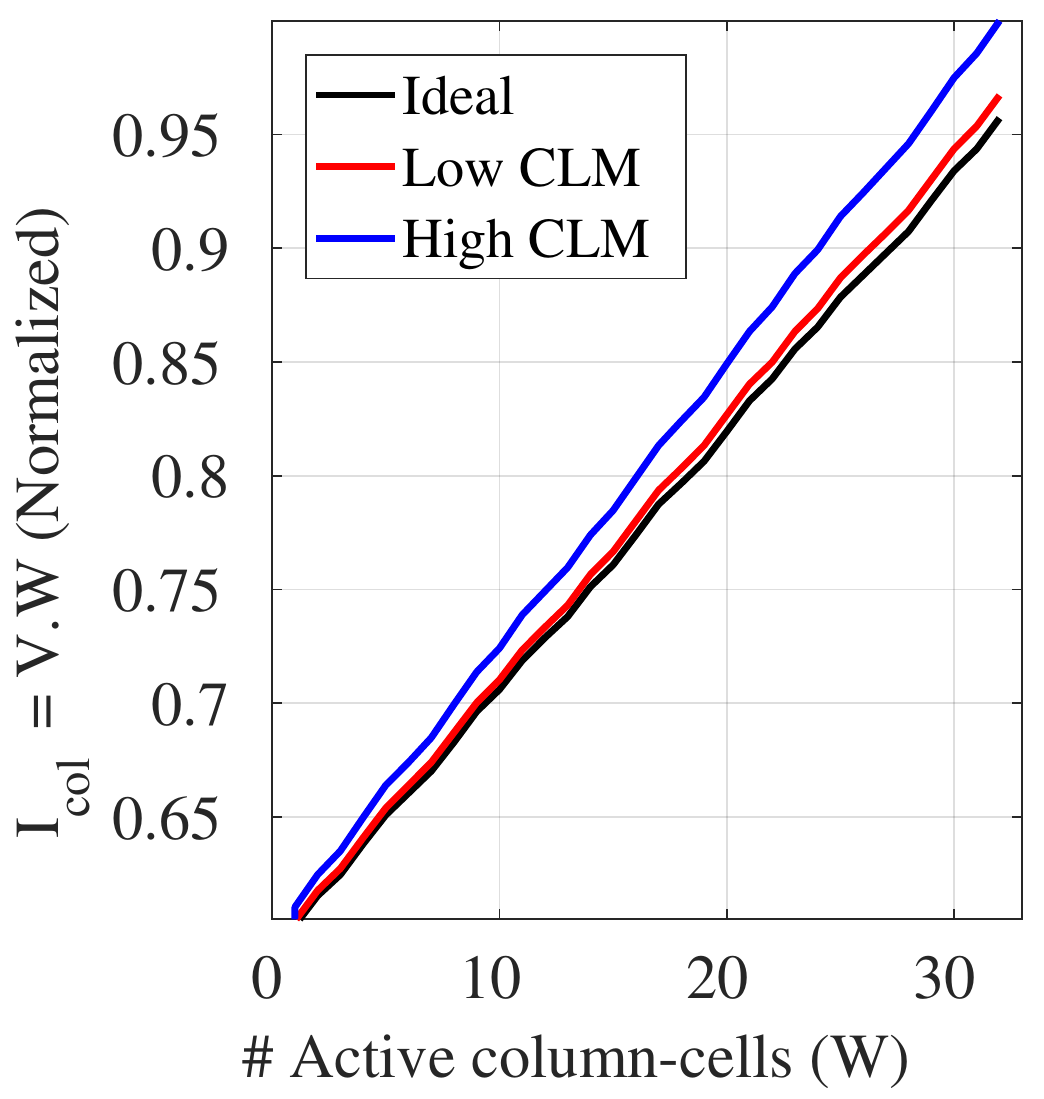}}
           \caption{}
        \end{subfigure}
        \begin{subfigure}[!t]{0.19\linewidth}
         \centering
        \begin{tikzpicture}[scale = 0.40]
        \pie{5/SRAM, 13/ DAC, 82/ADC}
        \end{tikzpicture}
        \vspace{2em}
        \caption{}
        \end{subfigure}
        \vspace{-0.25em}
    \caption{\small{(a) $V.W$ scalar product simulated in 45nm CMOS for density computation. (b) Effect of $V_{TH}$ variability in MC$^2$SRAM transistors to scalar product current $(\sigma(V_{TH})$ = 30 mV). (c) Current variability in MC$^2$SRAM controlled by DAC. (d) Effect of CLM in mirror transistors of DAC on scalar product accuracy. (e) Contribution of MC$^2$RAM peripherals to power consumption per sampling iteration.}}
    \label{fig:ckt_plots}
\end{figure*}

\begin{figure*}[!hbt]
    \centering
         \begin{subfigure}[!t]{0.35\linewidth}
        \centerline{\includegraphics[width=\linewidth, height=4cm]{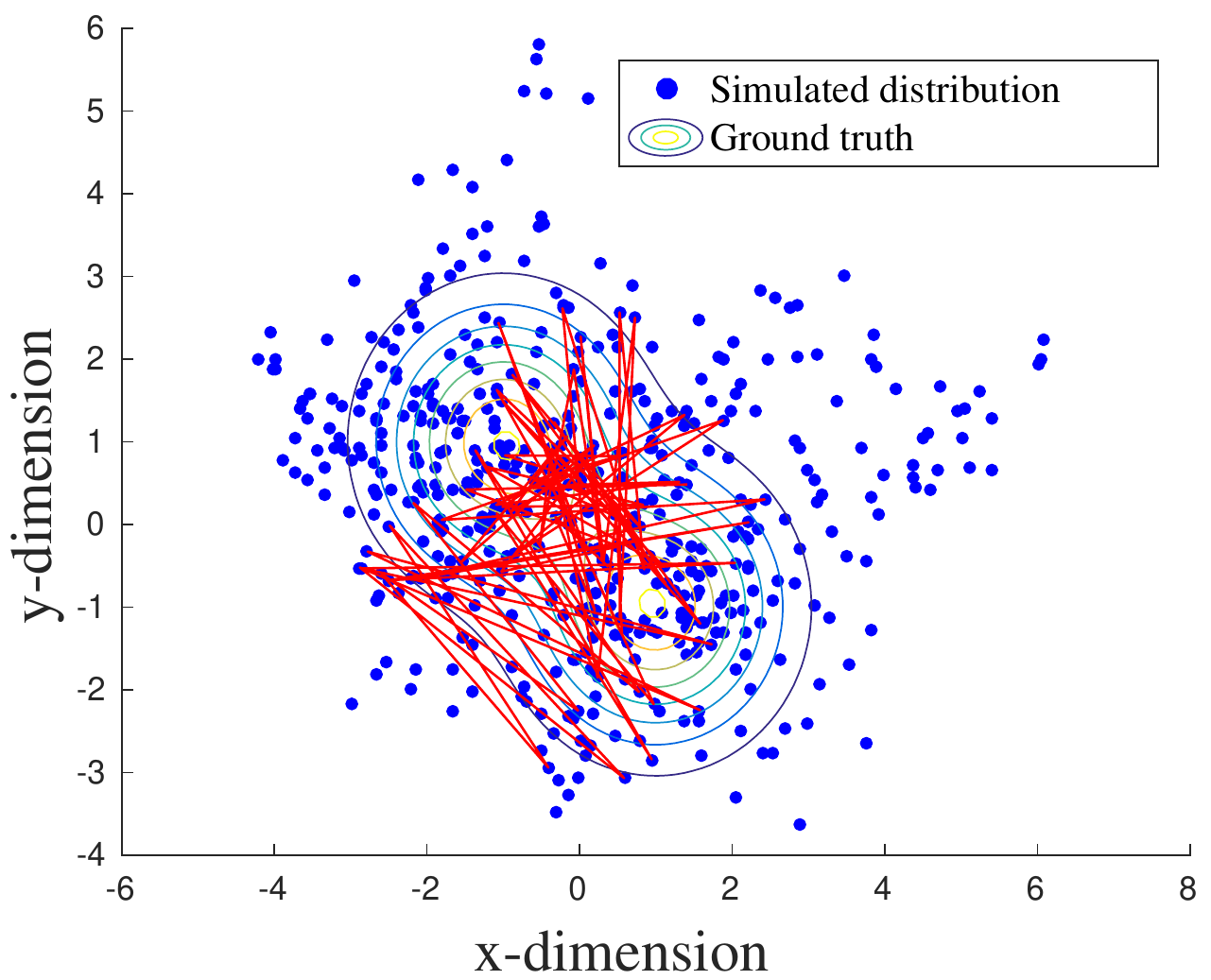}}
           \caption{Sampling within MC$^2$RAM}
        \end{subfigure}
         \begin{subfigure}[!t]{0.25\linewidth}
        \centerline{\includegraphics[width=\linewidth, height=4cm]{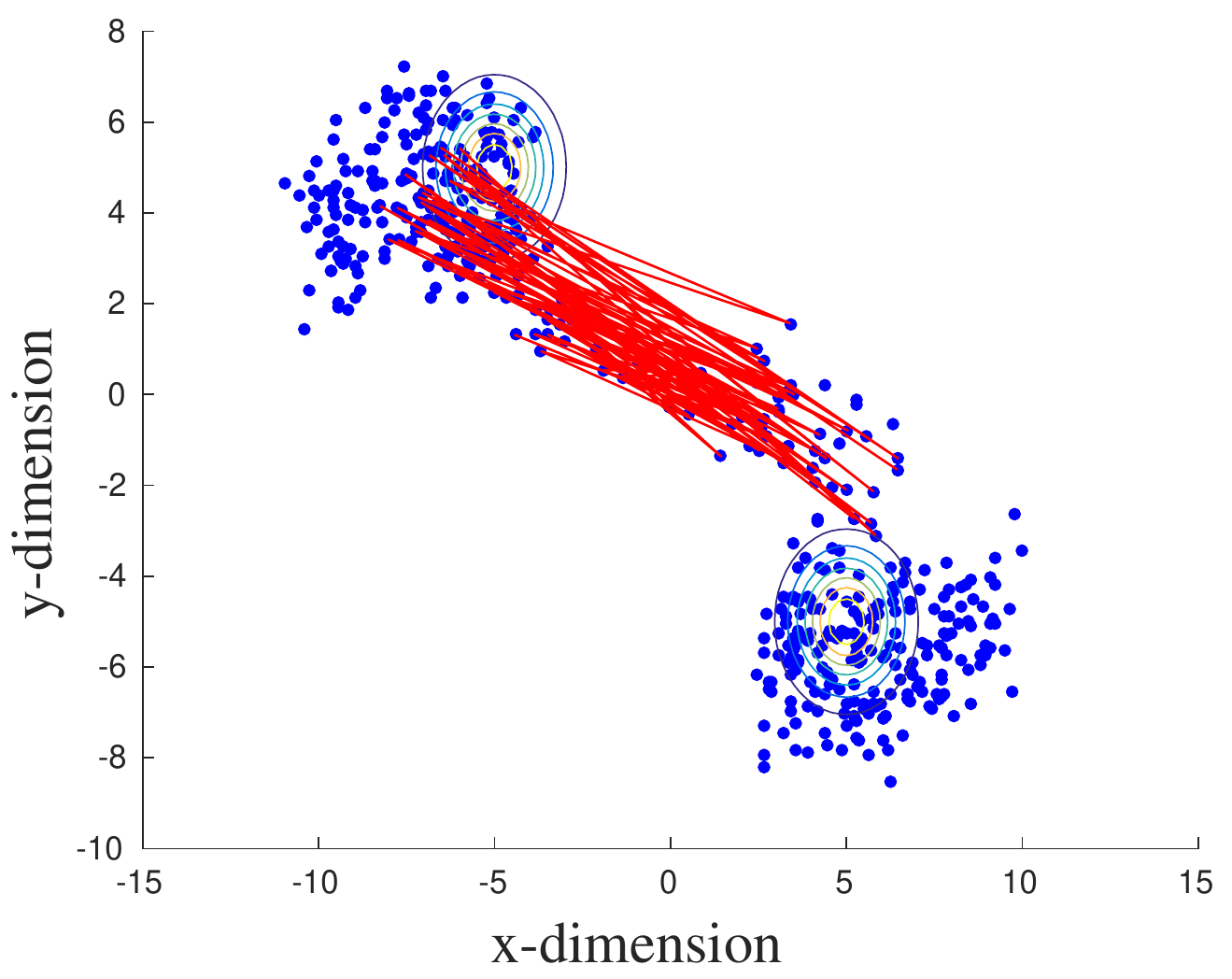}}
           \caption{Mean distance, d = 5}
        \end{subfigure}
        \begin{subfigure}[!t]{0.18\linewidth}
         \centerline{\includegraphics[width=\linewidth, height=4cm]{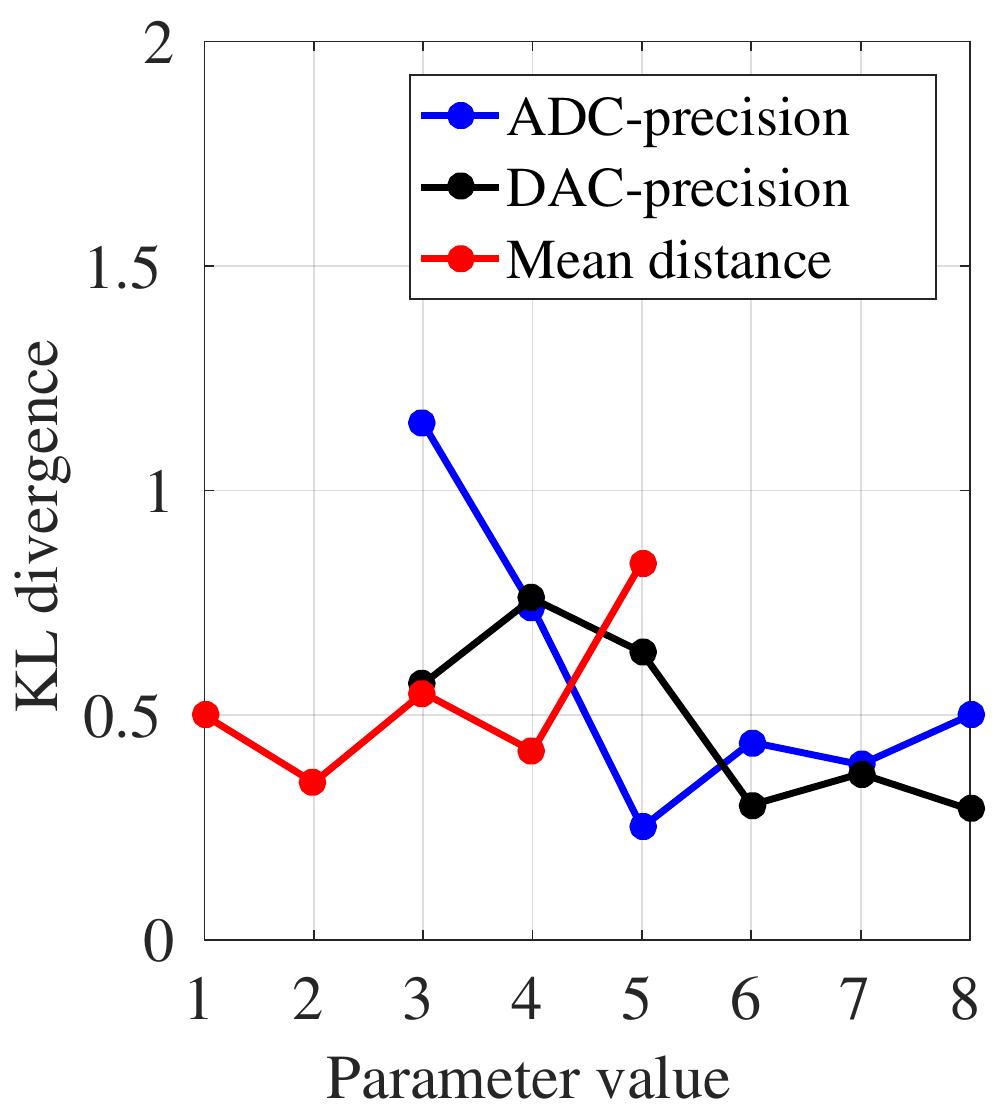}}
         \caption{}
        \end{subfigure}
         \begin{subfigure}[!t]{0.18\linewidth}
        \centerline{\includegraphics[width=\linewidth, height=4cm]{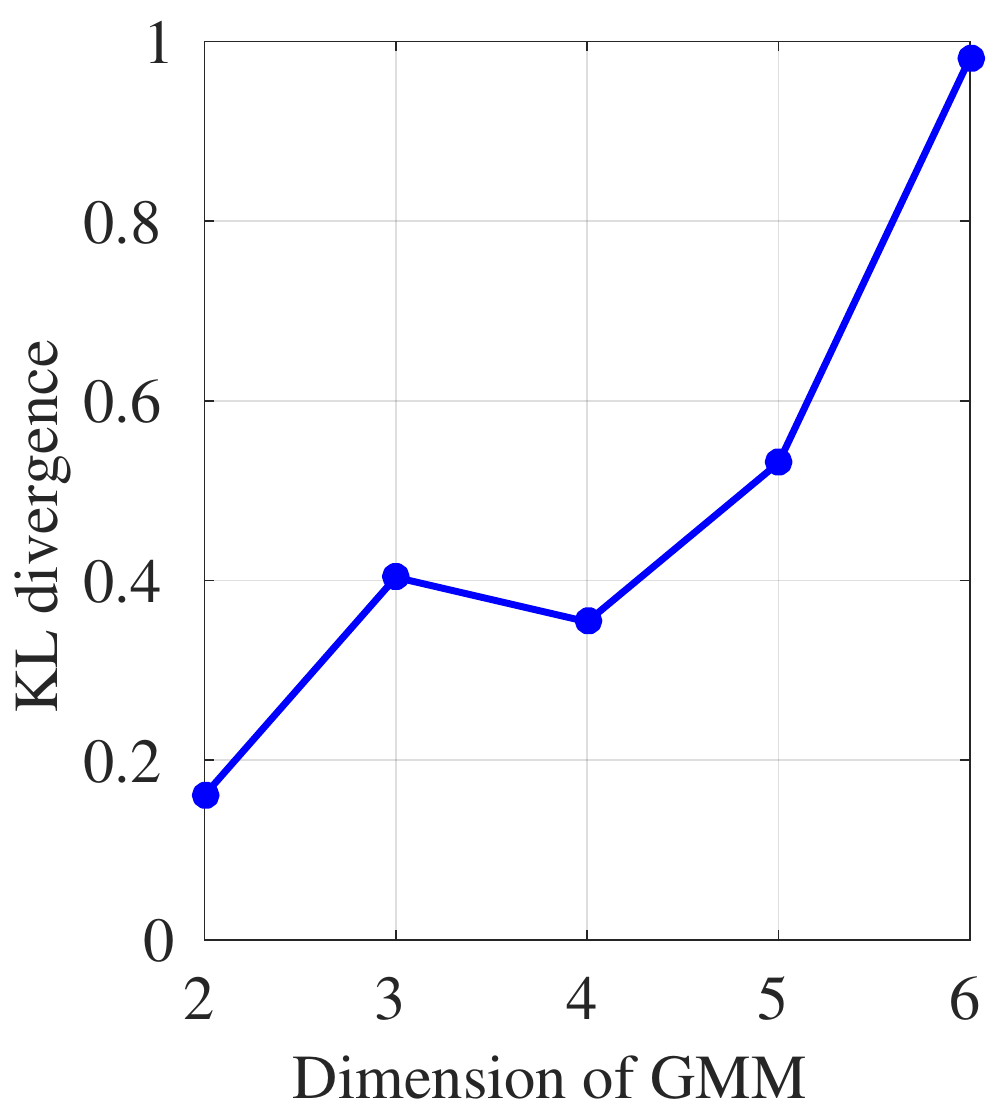}}
           \caption{}
        \end{subfigure}
        \vspace{-0.20em}
    \caption{\small{(a) Sampling distribution with 8-bit precision DAC, 6-bit precision ADC, VDD = 1V and mean distance = 1. (b) Sampling distribution with mean distance = 5. (c) KL divergence between ground truth (contour) and sampled distribution with sweeps performed on ADC bits-precision, DAC bits-precision and mean distance between GMM components. (d) KL divergence over a range of GMM dimensions.}}
    \label{fig:sampling}
    \vspace{-0.75em}
\end{figure*}

For the scalar product of two vectors $V$ and $W$, i.e., $V.W$, $W$ is stored in the 8-T SRAM cell-array and $V$ is copied to the DAC-operand buffer shown in Fig. \ref{fig:block_level_Ej}. An 8-T SRAM cell is shown in Fig. \ref{fig:circuits}(a) which has an additional scalar product port as shown in red in the figure. SRAM columns store $W$ in $n$-columns with $n$-bit precision. Digital-to-analog converter (DAC) converts the input $V$ into corresponding analog-mode current vector $I_v$ and applies to the product word line $WL_p$ of the cells. The basic approach for the scalar product is to use memory cells as current-mode AND gate. If an SRAM cell `j' stores bit `1', it allows the row DAC current $I_v$ to flow to its bit-line $BL_p$. The currents from each active cell in the column add up and follow $V.W$ as shown in Fig. \ref{fig:circuits}(c). The column multiplexer selects only one column at a time and the selected column-current will be read by an analog-to-digital converter (ADC) as shown in Fig. \ref{fig:circuits}(d) and (e) that converts column current to digital bits representing $V.W_i$, where $W_i$ is the $i^{th}$ precision binary vector of $W$. The column current is passed to an OP-AMP with a resistive feedback to convert the column current to corresponding analog voltage. The resistance value, $R$, is designed to match the operating range of ADC. OP-AMP in Fig. \ref{fig:circuits}(d) serves two purposes. It stabilizes the potential of the tail-end of the column, and it also biases the column tail potential to zero. The hold cell in the Fig. \ref{fig:circuits}(d) samples the output potential of OP-AMP and retains it after the OP-AMP is disconnected and bias-current of row DACs is turned off to save biasing power. The current of all n-columns are converted using ADC and combined with digital scaling to compute $V.W$. We use two-step flash ADCs \cite{maloberti1693488, Maloberti:2010:DC:1951885} that optimally balances area/power constraints without incurring excessive delay.

For the proposed implementation in 45nm CMOS, Fig. \ref{fig:ckt_plots}(a) shows HSPICE simulation for the scalar product using 32-row SRAM cell array matching the ideal. The current-mode processing in the proposed design gives significant advantages. The SRAM cells either act as current buffers or block the input current so that the variability in SRAM cell transistors has minimal impact to the accuracy of scalar product that posed challenge in \cite{Zhang2017InMemoryCO}. Also, the $V_{TH}$ variability of cell transistors does not affect the scalar product when DAC reference currents are sufficiently higher than SRAM leakage. Fig. \ref{fig:ckt_plots}(b-c) illustrate the variability analyses on the operation of SRAM. In Fig. \ref{fig:ckt_plots}(b), the column current follows a Log-Normal distribution when considering process variability due to SRAM transistors. In Fig. \ref{fig:ckt_plots}(c), with $\sigma(V_{TH})$ = 30 mV (black curve), the variation in column current is 1.43 (normalized against mean) that corresponds to the DAC reference current of 5 nA. Upon increasing the DAC current, the variability of column current reduces when normalized against the mean value. 

DAC in Fig. \ref{fig:circuits}(b) displays two critical non-idealities that affect the scalar product accuracy: (i) Channel length modulation (CLM) in the mirroring transistors that affects the scalar product accuracy posing dependence to \textit{W$L_p$} potential and (ii) non-ideal mirroring ratio due to process variability. We address CLM-induced precision degradation by reducing the turn-ON voltage of select switches in DAC to limit source-to-drain voltage of mirroring transistors, which improves the accuracy as shown in Fig. \ref{fig:ckt_plots}(d). To minimize process variability-induced non-ideal mirroring ratio in DAC, a set of calibrating transistors with small width \textit{$W_c$} relative to mirroring transistors are added to DAC in Fig. \ref{fig:circuits}(b). DAC mirror current is read against a reference to add \textit{$W_c$} until the current meets the desired level.

In MC$^2$RAM, storage of density function ($\mathcal{F}(x)$) parameters and sample generation $R$ is collocated within the same array by integrating RNG cells with SRAM cells. Since in a high-dimensional weight space many $R$ end up being rejected, co-locating the operations with the same SRAM array minimizes overheads and data movement. Fig. \ref{fig:circuits}(f) shows the RNG cell based on cross-coupled inverters \cite{mathew6339068}. The differential ends $Q$ and $Q_B$ are pre-charged to $V_{DD}$ when $CLK = 0$. When $CLK = 1$, the thermal noise resolves the meta-stability to generate a random bit. The random bits, stored in DAC operand buffer, can further be scaled with $\sigma^2$ within DAC as shown in Fig. \ref{fig:circuits}(g). The scaled $R/{\sigma^2}$ is used for density computation based on (4).
\vspace{-0.5em}

\section{Results and Discussions on Sampling}
We analyze simulated distribution with respect to the ground truth using scatter plots and KL divergence \cite{kldiv4218101}. The KL divergence between two discrete probability density functions $\mathcal{F}(x)$ and $\mathcal{G}(x)$ is measured as
\vspace{-0.5em}
\begin{equation}
D_{KL}(\mathcal{F}||\mathcal{G}) = \sum_{x}\mathcal{F}(x)log\Big(\frac{\mathcal{F}(x)}{\mathcal{G}(x)}\Big).
\end{equation}

We  considered a sample GMM, GMM$_T$, with $\mu = [1, -1;-1, 1]$, $\sigma = [1, 0; 0, 1]$, and $p = [0.5, 0.5]$. We also considered 500 samples to be sufficient to eliminate any statistical errors in KL divergence. We discard the first 50 samples as burn-in samples. For GMM$_T$, Fig. \ref{fig:sampling}(a) shows the sampling trajectory and distribution when the precision of DAC/ADC is 8 bits with 1 volts power supply. The contour lines in the figure represent ground truth GMM and the scattered dots in blue represent simulated distribution of samples from MC$^2$RAM. The trajectory of samples as shown in red in the figure corresponds to 75 random walks. Fig. \ref{fig:sampling}(b) illustrates sampling when mean distance parameter $d$ in $\mu = [d, -d; -d, d]$ between GMM components is set to 5. Fig. \ref{fig:sampling}(c) shows DAC/ADC imprecision tolerance limit in MC$^2$RAM which allows DAC and ADC to be low power/area. Sampling deviates from ground truth for ADC below 5-bit precision. Whereas, reduction in DAC precision has no significant impact in KL divergence. This also justifies our choice of using a low precision two-step flash ADC in MC$^2$RAM which has lower overhead for low to moderate precision design. Also, for fixed sampling iterations KL divergence is high for large separation between GMM components. As we go for higher dimensions, the deviation of samples from ground truth increases as shown in \ref{fig:sampling}(d). For a two-dimensional, two mixture GMM, the implementation consumes $\sim91\mu\/W$ power per sampling iteration and produces 500 samples in 2000 clock cycles on an average at 1 GHz clock frequency. The pie chart in Fig. \ref{fig:ckt_plots}(e) shows SRAM cells and DAC together contributing to 18\% of power consumption whereas the remaining 82\% is due to the ADC. The 10 comparators in the 2-step subranging flash ADC constitute 60\% of the power consumed by the ADC, however, the delay associated with flash ADC is 2 clock cycles, which improves sampling throughput in MC$^2$RAM.

\section{Conclusion}
We have presented a novel framework MC$^2$RAM that is a key to accelerate Markov chain Monte Carlo (MCMC) sampling for Bayesian Inference (BI). We exploit MC$^2$RAM to store parameters of posterior density of weights in BI and random number generation for high throughput Metropolis-Hastings (MH) based sample acceptance/rejection. The framework samples at low precision and power with tolerance to process variation thus removing latency/energy/safety bottlenecks associated with traditional von Neumann architecture that has spatially distant memory and processing elements.

\bibliographystyle{IEEEtran}
\bibliography{MC2RAM_ISCAS.bbl}

\end{document}